\documentclass{emulateapj}

\begin{document}


\title{WSRT Ultra-Deep Neutral Hydrogen Imaging of Galaxy Clusters at z$\approx$0.2, \\
 a Pilot Survey of Abell 963 and Abell 2192   }


\author{Marc Verheijen\altaffilmark{1}, J.H. van Gorkom\altaffilmark{2},
A. Szomoru\altaffilmark{3},K.S. Dwarakanath\altaffilmark{4}, 
B.M. Poggianti\altaffilmark{5}, and D. Schiminovich\altaffilmark{2}} 

\altaffiltext{1}{Kapteyn Astronomical Institute, University of
Groningen, PO Box 800, 9700 AV Groningen, The Netherlands;\\
verheyen@astro.rug.nl}
\altaffiltext{2}{Dept of Astronomy, Columbia University, 550 W 120th
Street, New York, NY 10027, USA; ds@astro.columbia.edu,\\
jvangork@astro.columbia.edu}
\altaffiltext{3}{Joint Institute for VLBI in Europe, Dwingeloo, The
Netherlands; szomoru@jive.nl}
\altaffiltext{4}{Raman Research Institute, Sadashivanagar, Bangalore
560 080, India; dwaraka@rri.res.in}
\altaffiltext{5}{INAF - Padova Astronomical Observatory, Vicolo
Osservatorio 5, 35122 Padova, Italy; poggianti@pd.astro.it}

\begin{abstract}

A pilot study with the powerful new backend of the Westerbork
Synthesis Radio Telescope (WSRT) of two galaxy clusters at z=0.2 has
revealed neutral hydrogen emission from 42 galaxies.  The WSRT probes
a total combined volume of 3.4$\times$10$^4$ Mpc$^3$ at resolutions of
54$\times$86 kpc$^2$ and 19.7 km/s, surveying both clusters and the
large scale structure in which they are embedded.  In Abell~963, a
dynamically relaxed, lensing Butcher-Oemler cluster with a high blue
fraction, most of the gas-rich galaxies are located between 1 and 3
Mpc in projection, northeast from the cluster core. Their velocities
are slightly redshifted with respect to the cluster, and this is
likely a background group.  None of the blue galaxies in the core of
Abell~963 are detected in H~I, although they have similar colors and
luminosities as the H~I detected galaxies in the cluster outskirts and
field. Abell~2192 is less massive and more diffuse. Here, the gas-rich
galaxies are more uniformly distributed. The detected H~I masses range
from 5$\times$10$^9$ to 4$\times$10$^{10}$ M$_{\odot}$. Some galaxies
are spatially resolved, providing rudimentary rotation curves useful
for detailed kinematic studies of galaxies in various environments.
This is a pilot for ultra-deep integrations down to H~I masses of
8$\times$10$^8$ M$_{\odot}$, providing a complete survey of the gas
content of galaxies at z=0.2, probing environments ranging from
cluster cores to voids.

\end{abstract}

\keywords{galaxy clusters: general --- galaxy clusters: individual(Abell 963,
Abell 2192)}

\section{Introduction}

It has long been recognized that the morphological mix of galaxy types
is very different in the centers of clusters than in the
field. Ellipticals and S0's dominate in dense clusters, spirals and
irregulars dominate elsewhere.  More recent studies have shown that
the galaxy population of clusters at intermediate redshifts evolves
over relatively short timescales. Beyond z$\approx$0.2, clusters have
a larger fraction of blue galaxies, indicative of ongoing star
formation; the so called Butcher-Oemler (B-O) effect (Butcher and
Oemler, 1978). The morphological mix in clusters also changes with
redshift. Although the fraction of ellipticals remains unchanged from
z=1 to 0, distant clusters have a significant fraction of spirals and
hardly any S0's, while this situation is reversed for more nearby
clusters (Dressler et al 1997, Fasano et al 2000, Lubin, Oke \&
Postman 2002). Furthermore, recent data from the local universe
suggests there are smooth gradients in star formation rate, gas
content and morphological mix, out to several Mpc from the cluster
centers (Goto et al 2003; Balogh et al 1998; Solanes et al 2001). More
specifically, there is an ongoing debate whether it is the cluster
environment that drives the morphological evolution of galaxies or
whether it is the accreted field population that evolves with
redshift. In a study of a classic B-O cluster at z=0.6, Tran et al
(2005) find that the blue starforming galaxies are predominantly
associated with an infalling structure, thus providing a strong
argument that the B-O effect is linked with infall.

Despite all these data, nothing much is known about the cold neutral
{\bf gas} content of galaxies beyond z=0.08. The gas content is a
critical parameter in environmentally driven galaxy evolution, since
cold gas forms the reservoir of fuel for star formation, and the
extended cold gas disks are sensitive tracers of tidal and
hydrodynamical interactions. The main impediments to observe H~I in
galaxies at larger redshifts are the necessarily long integration
times and man-made interference outside the protected 21cm band. So
far, H~I emission has been detected from only one galaxy at z=0.176
(Zwaan, van Dokkum \& Verheijen, 2001).  Nevertheless, if one were to
sample a large volume and observe many galaxies at once, it would be
worth the long integration time.

Radio synthesis telescopes, with their large field of view and
sufficient angular resolution to resolve individual galaxies, are the
ideal instruments for this. However, one drawback of synthesis
instruments has been the limited instantaneous velocity coverage,
typically no more than 2000 km/s, which is insufficient to cover the
range of velocities usually seen in clusters (up to 5000 km/s). This
has now changed with the new backend of the WSRT. In a single
pointing, 18,000 km/s can be covered with sufficient velocity
resolution, probing not only the entire velocity range of the cluster
but also a very significant volume in front of and behind the cluster.
Such an observation would provide the gas content of all galaxies in
the cluster and the surrounding large scale structure.  Here, we
report on first results from a pilot study for just such a survey.

Two clusters were selected, known to be very different in their
dynamical state and star formation properties.  {\bf Abell 963}, at
z=0.206, has a velocity dispersion of 1350 km/s and is one of the
nearest B-O clusters with a high fraction (19\%) of blue galaxies
(Butcher et al 1983).  This lensing X-ray cluster is unusually relaxed
with less than 5\% substructure (Smith et al 2005).  {\bf Abell 2192},
at z=0.188, has a velocity dispersion of 650 km/s and is much more
diffuse. So far, it has not been detected in X-rays and the fraction
of blue galaxies in this cluster has not yet been determined. Both
clusters are observed by the Sloan Digital Sky Survey (SDSS).  In
Figure 1 we show a pie-diagram taken from the SDSS along a great
circle passing through both clusters. The boxes indicate the volumes
probed by the WSRT. Clearly, the clusters only occupy a tiny fraction
of the volume.  We also obtained deeper B- and R-band images for both
clusters (Figure 2, Plate 1) and some optical redshifts for Abell~2192
with Hydra on WIYN at Kitt Peak. The box in Figure 2 indicates the
area of A963 observed by Butcher et al (1983).

\begin{figure}[t]
\epsscale{1.15}
\figurenum{1}
\plotone{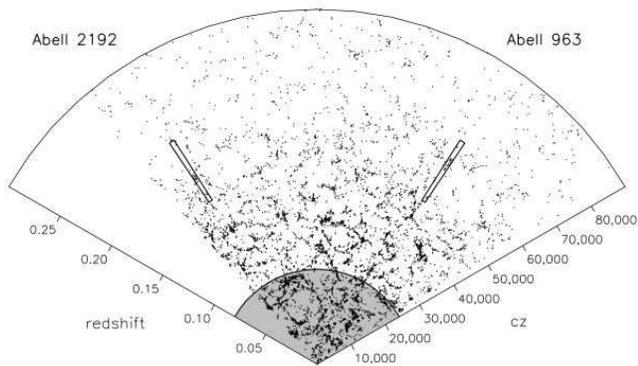}
\caption{SDSS pie-diagram along a great circle passing through both
clusters studied here. The grey area indicates the local volume in
which clusters have been studied in H~I so far. Boxes indicate the volumes
surveyed by us in H~I with the WSRT.}
\label{fig1}
\end{figure}

\section{Observations}

To prove the feasibility of observing the H~I content of galaxies in
and around these clusters with the WSRT, Abell 963 was observed for
20$\times$12 hrs in February 2005, and Abell 2192 was observed for
15$\times$12 hrs in July 2005. Each cluster was observed with a single
pointing. The correlator was configured to give eight, partially
overlapping, 10 MHz bands (IVC's) with 256 channels per band and 2
polarizations per channel. The frequency range covered is 1220$-$1160
MHz or 0.164$<$z$<$0.224. After standard data reduction procedures,
the data from all 8 IVCs were combined for each cluster into a single
datacube comprising 1600 channel maps. The achieved typical rms noise
levels per frequency channel are 68 $\mu$Jy/beam at 1178 MHz for A963,
and 91 $\mu$Jy/beam at 1196 MHz for A2192. After Hanning smoothing,
the restframe velocity resolution is 19.7 km/s while the synthesized
beam is 17$\times$27 arcsec$^2$ or 54$\times$86 kpc$^2$ at 1190 MHz
($\Omega_{\rm M}$=0.27, $\Omega_{\rm \Lambda}$=0.73, H$_0$=71
km/s/Mpc).

In search of H~I emission, we further smoothed the data to a velocity
resolution of 40 km/s and visually inspected both datacubes. Based
solely on the H~I datacubes, we identified 20 H~I detections in the
field of A963, and 30 in that of A2192. Subsequently, we queried the
SDSS images for optical counterparts within the synthesized beam and
found 19 objects for A963 and 23 for A2192. We consider these 42 H~I
detections as secure and the remaining 8 as tentative, given the
shallowness of the SDSS. Unfortunately, SDSS spectroscopy, as an
additional check, is only available for one of our H~I detections.
Assuming that the surveyed volume (Figure 1) is representative of the
local Universe at large, in terms of galaxy density distributions,
allows us to compare our H~I detection rate with those obtained in
blind H~I surveys at lower redshifts.  Given the local H~I mass
function (Zwaan et al 2003), the surveyed volume, the achieved noise
levels, and an equivalent detection threshold of 4 sigma in 3
resolution elements, the predicted number of detectable galaxies is 22
for A963 and 17 for A2192.  The detection rate in our pilot data is
roughly as expected.

\begin{figure}[t]
\epsscale{1.15}
\figurenum{3}
\plotone{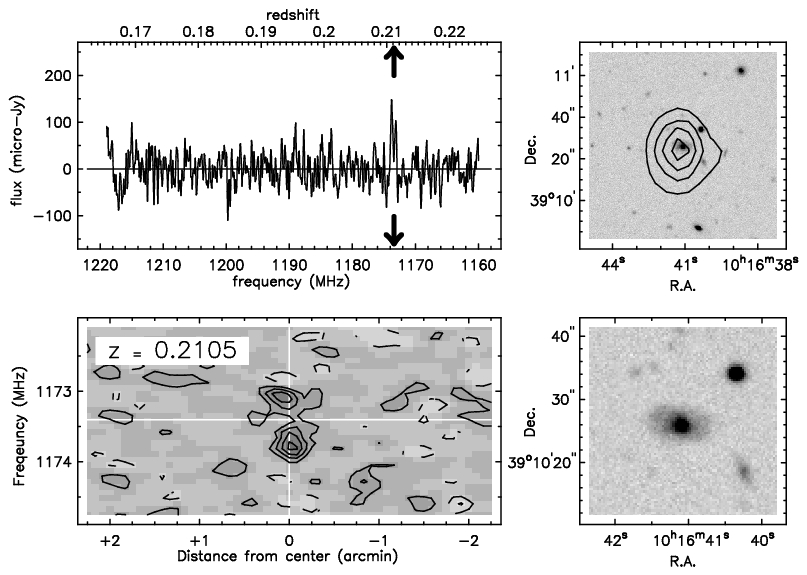}
\caption{One example of our H~I detections; a spatially resolved
galaxy at the same redshift as Abell~963.  Upper left: HI spectrum
over the full velocity range. Arrows indicate the redshift (top) and
observed frequency (bottom) of the H~I emission. Lower left:
position-velocity diagram extracted from the H~I datacube, taken along
the major axis of the galaxy. The horizontal white line indicates the
systemic velocity. The vertical white line coincides with the spatial
center of the galaxy. Upper right: integrated H~I map in
countours overlayed on an R-band image in grayscales. Lower right:
blow-up of the optical image, showing the morphology of a
normal spiral galaxy.}
\label{fig3}
\end{figure}

\section{Results}

The quality of the data is exquisite. In Figure 3 we show typical data
on an individual galaxy in the outskirts of Abell~963. The upper panel
shows the double peaked profile.  Although the total H~I image, shown
in the bottom left panel, seems barely resolved, the position velocity
diagram taken along the photometric major axis shows that the
kinematics helps to spatially resolve the galaxy.  Its rudimentary
rotation curve looks promising for future Tully-Fisher studies of
galaxy samples in different environments at this redshift.  Individual
H~I detections have H~I masses between 5$\times$10$^9$ M$_{\odot}$,
close to our detection limit, and 4$\times$10$^{10}$ M$_{\odot}$.

\begin{figure}[t]
\epsscale{1.15}
\figurenum{4}
\plotone{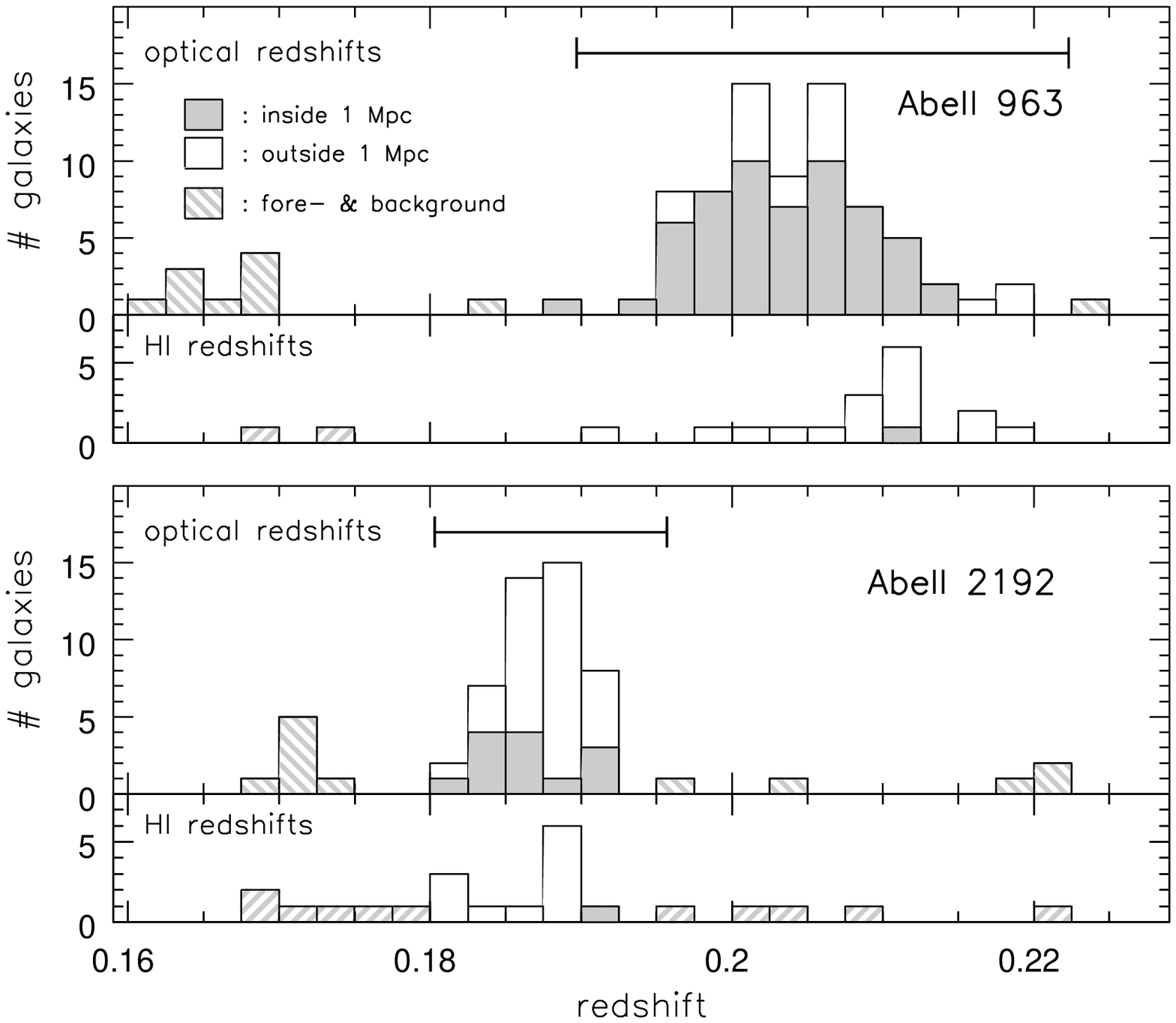}
\caption{Redshift distributions of optically selected galaxies with
optical redshifts in the upper panels, compared to those of H~I
detected galaxies in the lower panels. The adopted cluster redshift
ranges are indicated by horizontal bars. Dashed histograms indicate
fore- and background galaxies. Solid histograms indicate galaxies
within 1 Mpc from the cluster centers, open histograms indicate
galaxies outside 1 Mpc.}
\label{fig4}
\end{figure}

In Figure 4 we show the redshift distributions of the H~I detected
galaxies compared to those of optically selected galaxies with optical
redshifts. We identify the clusters with a velocity range of $\pm$
3$\sigma$ around the cluster mean velocity, where $\sigma$ is the
cluster velocity dispersion. This corresponds to $0.1897<z<0.2223$ for
A963 and $0.1803<z<0.1957$ for A2192, as indicated by the horizontal
bars. For A2192, half of the H~I detections are outside this velocity
range (dashed histogram) and are located at large projected distances
from the cluster center. These are likely fore- and background
galaxies.  For A963, most of the H~I detections are within the cluster
velocity range but as we shall discuss below, many of those are likely
to be background galaxies. Galaxies that are located within a
projected distance of 1 Mpc from the cluster centers are indicated
with filled histograms and those outside 1 Mpc with open
histograms. Cluster membership obviously depends on both, with an
expected decreasing velocity range for cluster members at larger
projected distances.

Figure 5 shows the distributions of galaxies on the sky with known
redshifts within the velocity ranges of the clusters. Solid symbols
indicate our H~I detections while open symbols indicate optically
selected galaxies with optical redshifts. To give a sense of scale,
the small dotted circles have a radius of 1 Mpc at the distance of the
clusters.  The large dashed circles indicate the FWHM of the primary
beam of the WSRT. Note that R$_{200}$ is 3 Mpc and 1.5 Mpc for
Abell~963 and Abell~2192 respectively. The field of view of the WSRT
extends far beyond the clusters.  Although the H~I detection rate is
quite similar for both regions, the spatial distributions of the H~I
detections are very different for the two clusters.  The H~I detected
galaxies in the field of A963 are strongly clustered toward the
northeast of the cluster, while the H~I detected galaxies in A2192
appear to be more uniformly distributed over the surveyed area. For
each cluster we only detect one galaxy within 1 Mpc from the
center. The galaxies to the NE of A963 are also strongly clustered in
velocity, centered near z=0.21 in Figure 4.  Hence their mean velocity
is redshifted by about 1200 km/s with respect to the cluster
mean. Their mean projected distance from the cluster center is less
than 2 Mpc. There is a velocity gradient from east to west of 2100
km/sec over approximately 4 Mpc at the distance of the cluster. This
could be part of the filament seen in Figure 1 and is likely to be in
the background to Abell~963.

\begin{figure}[t]
\epsscale{1.15}
\figurenum{5}
\plotone{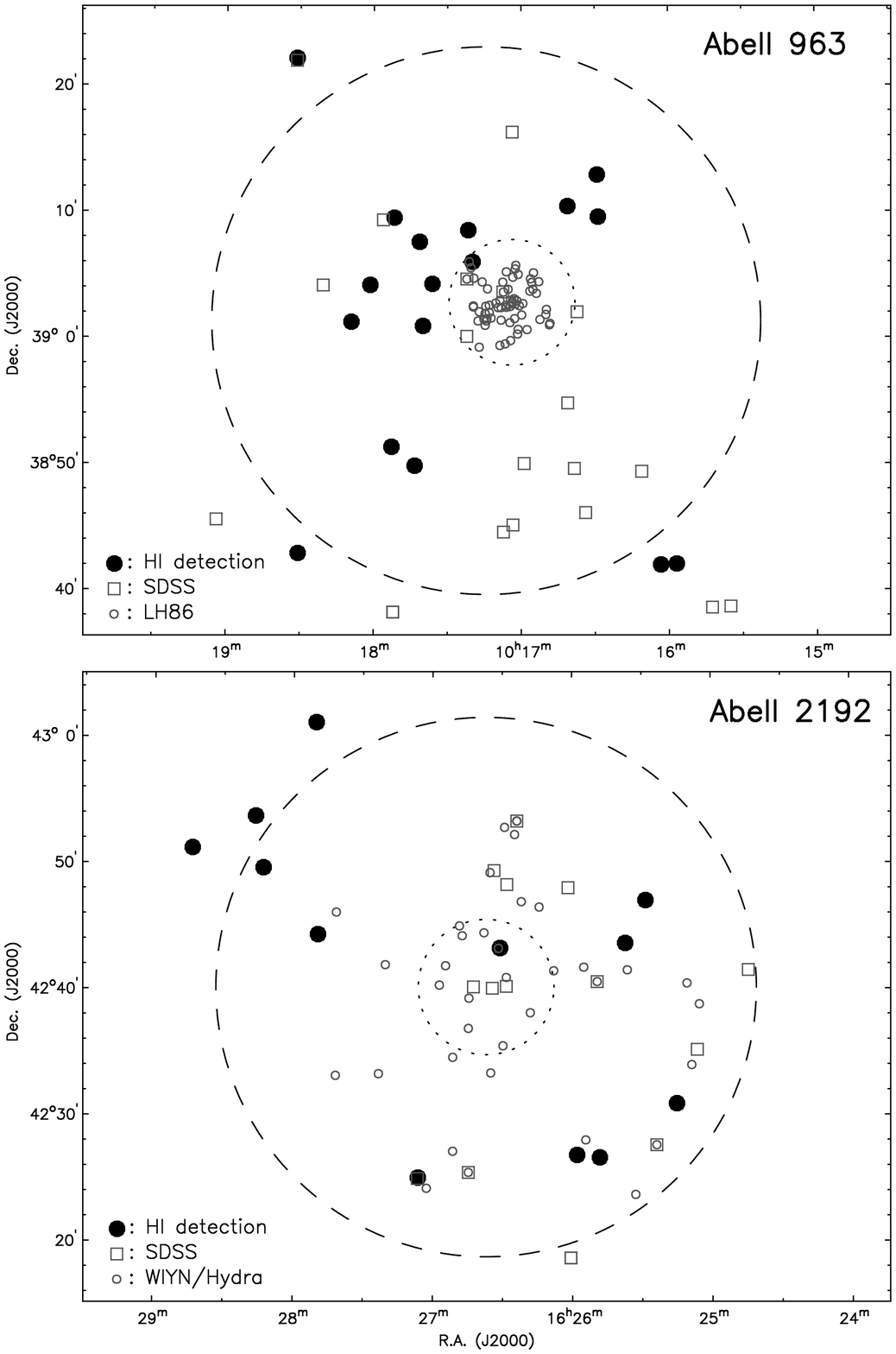}
\caption{Sky distributions of galaxies with optical and H~I redshifts
within the velocity ranges of the clusters. Solid symbols: H~I detected
galaxies. Open squares: galaxies with optical SDSS redshifts. Open
circles: galaxies with optical redshifts from Lavery \& Henry (1986,
LH86) for A963, and from our WIYN/Hydra spectroscopy for A2192.  Small
dotted circles have a radius of 1 Mpc at the distance of the clusters.
Large dashed circles indicate the FWHM of the primary beam of the
WSRT.}
\label{fig5}
\end{figure}

We can ask whether the galaxies detected in HI differ in other ways
from the non-detected galaxies. In Figure 6 we show, with identical
symbols as in Figure 5, a combined color-magnitude diagram for both
clusters with all the galaxies within their redshift ranges. Model
magnitudes from SDSS DR5 are used, corrected for Galactic extinction
and k-corrected following Blanton \& Roweis (2007). Absolute magnitudes
are obtained by subtracting an effective distance modulus
corresponding to the mean redshifts of the clusters. This figure
clearly shows the red sequence characteristic of a cluster population
as well as the 'blue cloud'. The main conclusion we can draw from
Figure 6 is that our H~I detected galaxies are in the same magnitude
and color range as the non-detected blue galaxies within 1 Mpc from
the center of A963 (most of the small circles in the blue cloud). This
suggests that it is the location of the blue galaxies that determines
the H~I detection rate.

This result gets enhanced when we probe deeper by stacking the H~I
spectra of galaxies with known optical redshifts. We have done this
separately for 12 blue galaxies within 1 Mpc of A963 with optical
redshifts from Lavery and Henry (1986), and for 14 blue galaxies
without an H~I detection outside 1 Mpc in both fields. In Figure 7 we
show the spectra for both samples (thick lines). As a control
experiment we also show stacked spectra at 8 positions offset from the
galaxies, but shifted to the same velocity (thin lines).  There is no
statistical detection of the blue galaxies within 1 Mpc of A963. In
contrast, the blue galaxies in the fields outside the central Mpc do
show a statistical detection in the stacked H~I spectra.

\begin{figure}[t]
\epsscale{1.15}
\figurenum{6}
\plotone{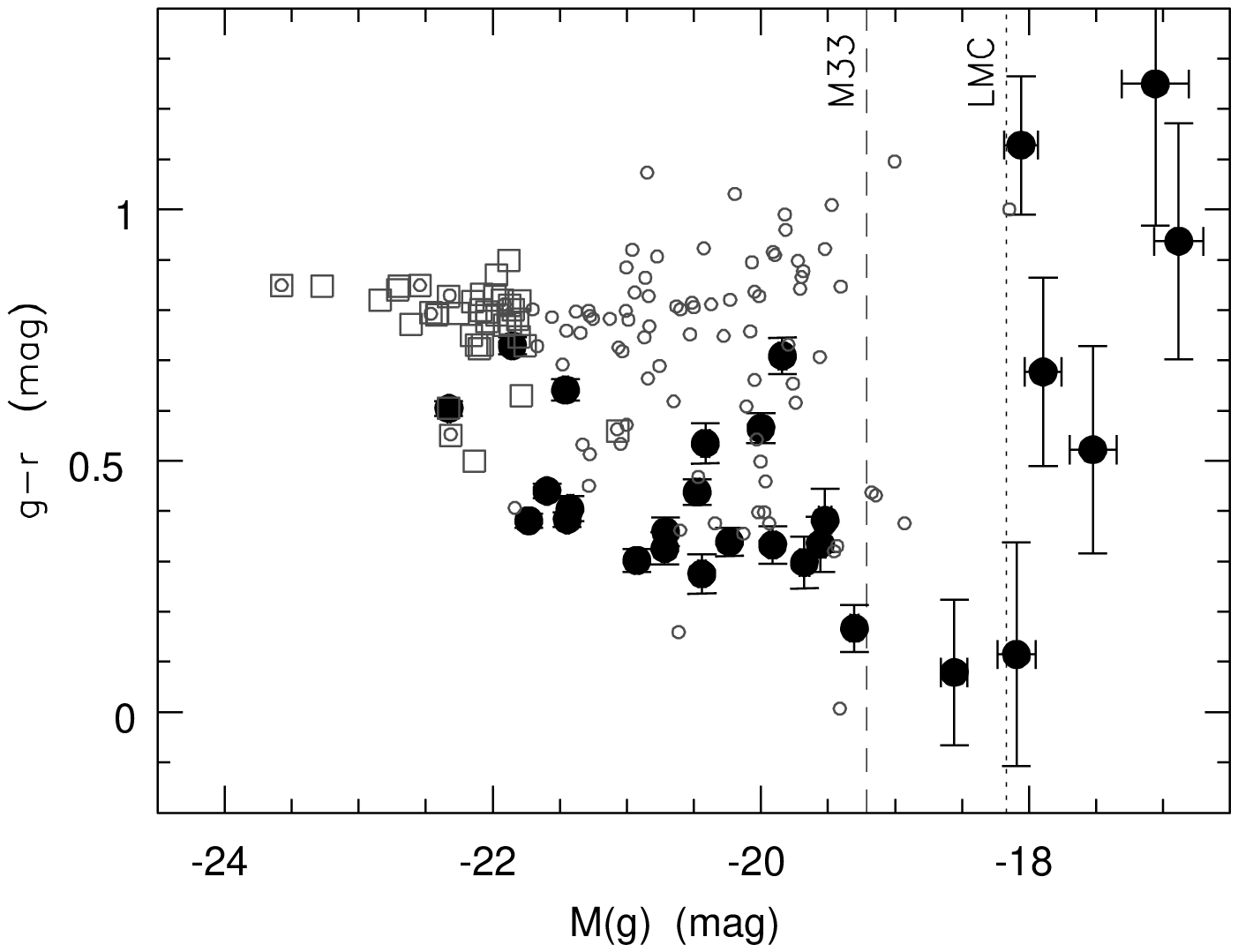}
\caption{SDSS-based color-magnitude diagram for both clusters
combined, including only galaxies with optical or H~I redshifts within
the cluster redshift ranges (see Fig.~4). Model magnitudes from SDSS
DR5 are used, corrected for Galactic extinction and k-corrected
following Blanton \& Roweis (2007), but without internal extinction
corrections. Absolute magnitudes are obtained by subtracting an
effective distance modulus corresponding to the mean redshifts of the
clusters. Symbols are identical to those in Figure~5.  Dashed vertical
lines indicate the magnitudes of M33 and the LMC.}
\label{fig6}
\end{figure}

\section{Discussion}

We present the results of a pilot study to demonstrate the feasibility
of performing an H~I emission line survey at z=0.2. Our detection rate
and achieved noise levels show that it is now entirely feasible to do
deep searches for H~I in emission at cosmologically interesting
distances and even obtain spatially resolved kinematics for individual
galaxies.  The preliminary results have interesting implications.
While our detection rate is similar for both clusters, the
distribution of the H~I detections is more clustered for the dense and
dynamically relaxed B-O cluster. The most noticable result is that
within the redshift range of A963 we detect 16 very gas-rich galaxies
outside a radius of 1 Mpc from the cluster center, and only one just
within that radius. At the same time we know that A963 has a
significant fraction of blue galaxies in that central cluster
region. Our stacked H~I spectra suggest that the central blue galaxies
have significantly smaller H~I masses (on average) than similar blue
galaxies outside the central Mpc.  Since the H~I detected galaxies are
of similar color and magnitude as the non-detected blue galaxies in
the central region, we conclude that it is the location that matters.
The central blue galaxies seem to have lost (a significant fraction
of) their gas when they came to within 1 Mpc of the center of A963.

Our final survey will go down to an H~I mass limit of 8$\times$10$^8$
M$_{\odot}$. We expect to detect hundreds of galaxies in the clusters
and in the large scale structure in which they are embedded.  The
survey goes out to 4 Mpc from the cluster centers in the plane of the
sky (Figure 5), and covers a velocity range of 18,000 km/s (Figure 4),
including filaments and voids (Figure 1). This will give the first
optically unbiased H~I survey at z=0.2 with enough sensitivity to
detect even galaxies like the LMC in a volume similar to that of the
entire local Universe out to 25 Mpc.

\begin{figure}[b]
\epsscale{1.15}
\figurenum{7}
\plotone{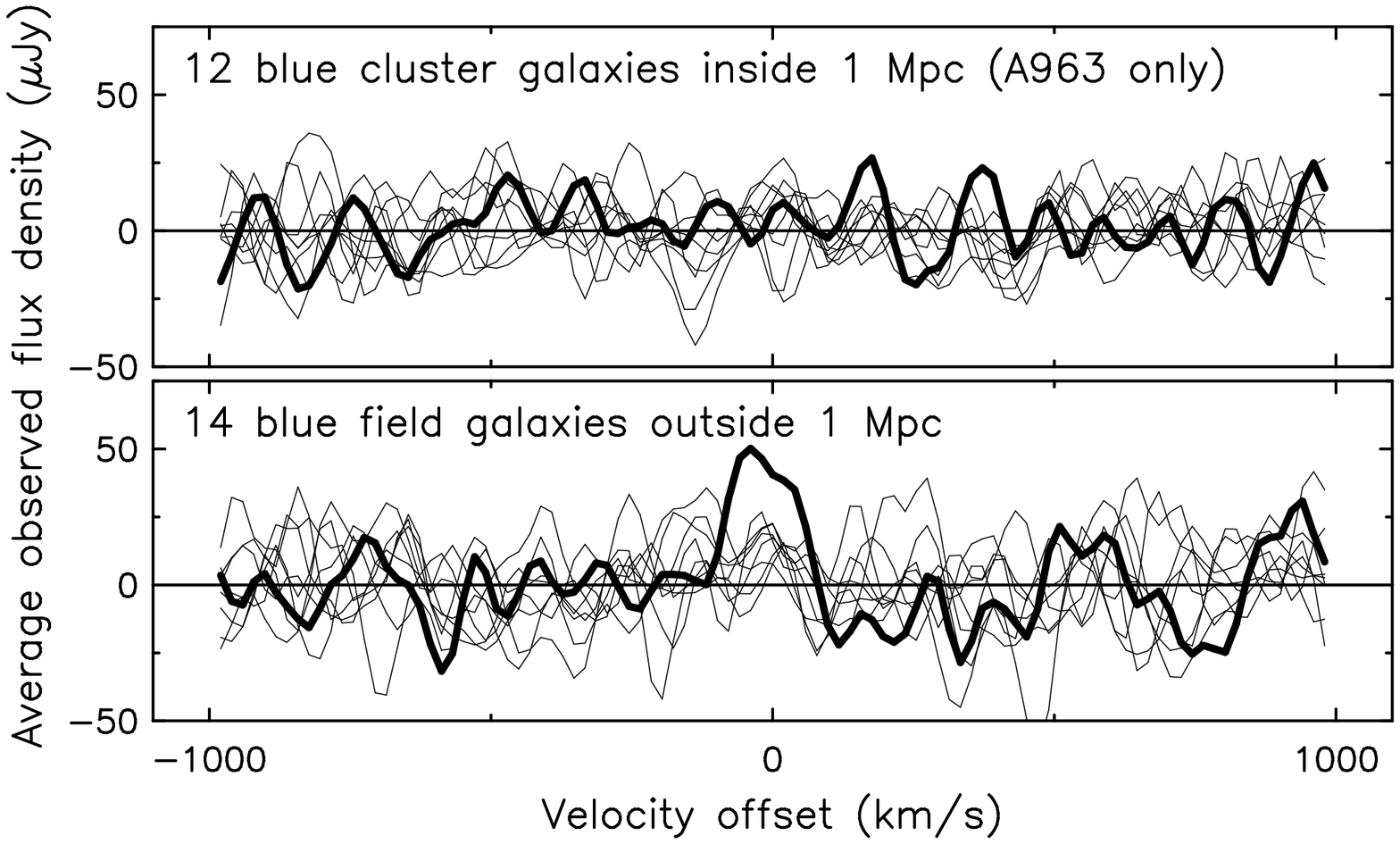}
\caption{Stacked H~I spectra of blue galaxies with optical redshifts,
not detected individually in H~I. Two different galaxy populations are
considered.  Top: blue B-O galaxies within 1 Mpc of the core of A963;
bottom: blue galaxies in the field of A2192 outside 1 Mpc of the
cluster core. Thin black lines indicate stacked spectra from 8 
spatially offset positions.}
\label{fig7}
\end{figure}

\acknowledgments 
We thank R. Lavery for providing redshifts of galaxies in A963,
T.~Oosterloo and K.~Kova\v{c} for providing the optical images of
A963, and S.~Trager for useful discussions.  The WSRT is operated by
the Netherlands Foundation for Research in Astronomy with support from
the Netherlands Foundation for Scientific Research. The WIYN
Observatory is a joint facility of the Universities of Wisconsin,
Indiana and Yale, and the NOAO. This work benefitted from a NSF grant
AST-06-07643 to Columbia University.  This work used the Sloan Digital
Sky Survey Archive. Its full acknowledgement is found at
http://www.sdss.org.

\clearpage

\begin{figure}[t]
\epsscale{1.15}
\figurenum{2}
%\plotone{figure2.eps}
\caption{Plate (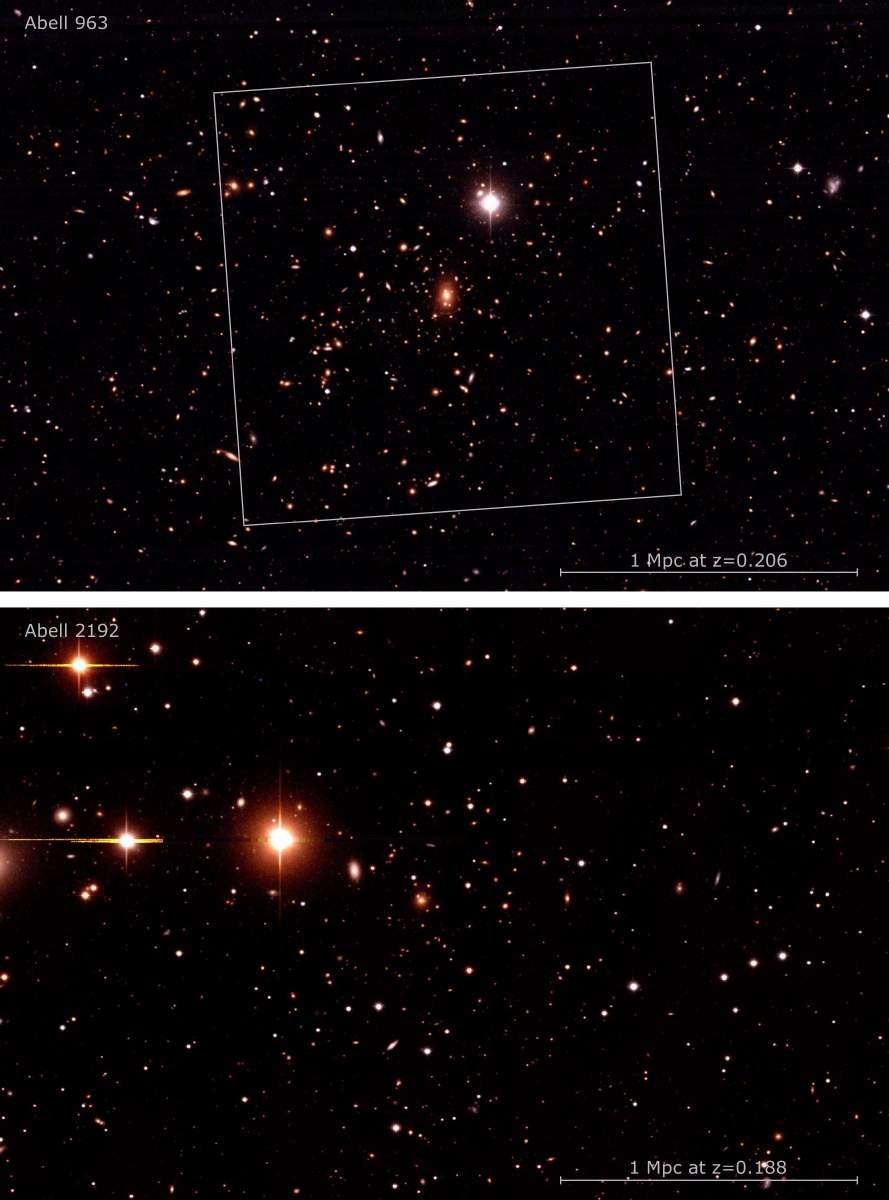) Optical images, 1.5$\times$2 Mpc on a
side, of the central regions of Abell 963 (top) and Abell 2192
(bottom). Colors are constructed from B- and R-band images taken with
the Wide Field Camera on the Isaac Newton Telescope at La Palma for
Abell 963, and with the NOAO Mosaic Camera on the 0.9m at Kitt Peak
for Abell 2192. The color scales are slightly different for the two
images. The area of A963 within which Butcher et al (1983) determined
the fraction of blue galaxies, is indicated with a box.}
\label{fig2}
\end{figure}


\begin{thebibliography}{}
\bibitem[Balogh et al 1978]{bal78} Balogh, M.L., Schade, D., Morris, S., 
Yee, H.K.C., Carlberg, R.G., \& Ellingson, E. 1998, \apjl, 504, L75  
\bibitem[Blanton and Roweis (2007)]{br07} Blanton, M.R. \& Roweis,
S. 2007, \aj, 133, 734
\bibitem[Butcher et al (1983))]{but83} Butcher, H., Wells, D.S. \&
Oemler, A. Jr. 1983, \apjs, 52, 183
\bibitem[Butcher and Oemler(1978)]{bo78} Butcher, H. \&
Oemler, A. Jr.   1978, \apj, 219, 18
\bibitem[Dressler et al (1997)]{dre97} Dressler, A. et al 1997, \apj, 490, 577
\bibitem[Fasano et al (2000)]{fas00} Fasano, G., Poggianti, B.M., 
Couch, W.J., Bettoni, D., Kjaergaard, P., \& Moles, M. 2000, \apj, 542, 673
\bibitem[Goto et al (2003)]{got03} Goto, T., et al 2003, \mnras, 346, 601
\bibitem[Kodama et al 2001]{kod01} Kodama, T, Smail, I., Nakata, F., 
Okamura, S., \& Bower, R.G. 2001, \apjl, 562, L9
\bibitem[Lavery and Henry 1986]{lh86} Lavery, R.J. \& Henry, J.P. 1986, \apjl,
304, L5
\bibitem[Lubin et al (2002]{lop02} Lubin, L.M., Oke, J.B., \& Postman, M. 
2002 \aj, 124, 1905 
\bibitem[Smith et al (2005)]{smi05} Smith, G.P., Kneib, J.-P., Smail, I., 
Mazzotta, P., Ebeling, H., \& Czoske, O. 2005, \mnras, 359, 417
\bibitem[Solanes et al (2001)]{sol01} Solanes, J.M., Manrique, A., 
Garcia-Gomez, C., Gonzalez-Casado, G., Giovanelli, R., \& Haynes, M. 2001, 
\apj, 548, 97  
\bibitem [Tran et al (2005)]{tra05} Tran, K.-V., van Dokkum, P., 
Illingworth, G.D., Kelson, D., Gonzalez, A., \& Franx, M. 2005, \apj, 619, 134 
\bibitem [Zwaan et al (2001)]{zdv01} Zwaan, M., van Dokkum, P. \& Verheijen,
M 2001, Science, 293, 1800
\bibitem[Zwaan et al (2003)]{zwa03} Zwaan, M. et al 2003, \aj, 125, 2842
\end{thebibliography}
\end{document}